\documentclass{osa-article}

\journal{oe}
\articletype{Research Article}
\usepackage{tikz}
\usepackage{subcaption}

\begin{document}

% \title{Topologically induced oblique wave propagation in photonic fiber arrays }
% \title{Oblique wave propagation in topological photonic crystal fiber arrays}
\title{Hybrid topological guiding mechanisms for \newline photonic crystal fibers}
% \title{Topological light-trapping along photonic \newline crystal fibers}
% \title{Localisation via topological effects in photonic crystal fiber arrays}

\author{Mehul Makwana,\authormark{1,2*} Richard Wiltshaw,\authormark{1}, S\'ebastien Guenneau\authormark{3} and Richard Craster,\authormark{1,3,4}}

\address{\authormark{1}Department of Mathematics, Imperial College London, London SW7 2AZ, UK\\
\authormark{2}Multiwave Technologies AG, 3 Chemin du Pr\^e Fleuri, 1228, Geneva, Switzerland\\
\authormark{3}UMI 2004 Abraham de Moivre-CNRS, Imperial College London, London SW7 2AZ, UK\\
\authormark{4}Department of Mechanical Engineering, Imperial College London, London SW7 2AZ, UK
}

\email{\authormark{*}mehul.makwana07@imperial.ac.uk} %% email address is required



\begin{abstract}
We create hybrid topological-photonic localisation of light by introducing concepts from the field of topological matter to that of photonic crystal fiber arrays. S-polarized obliquely propagating electromagnetic waves are guided by hexagonal, and square, lattice topological systems along an array of infinitely conducting fibers. 
The theory utilises perfectly periodic arrays that, in frequency space, have gapped Dirac cones producing band gaps  demarcated by pronounced valleys locally imbued with a nonzero local topological quantity. These broken symmetry-induced stop-bands allow for localised guidance of electromagnetic edge-waves along the crystal fiber axis. Finite element simulations, complemented by asymptotic techniques, demonstrate the effectiveness of the proposed designs for localising energy in finite arrays in a robust manner. %, even in the absence of an infinite conducting outer  boundary. %Potential applications of our results reside in the generation of topological photonic and phononic crystal fibers for telecommunications.
\end{abstract}
%%%%%%%%%%%%%%%%%%%%%%%%%%  body  %%%%%%%%%%%%%%%%%%%%%%%%%%

\section{Introduction}

Photonic crystal fibers (PCFs) guide light by corralling it within a periodic array of microscopic air holes, or dielectric inclusions, that run along the entire fiber length \cite{Russell_2003,knight_2003,pcfbook_2003,zolla_foundations_2005}. Largely through their ability to overcome the limitations of conventional fiber optics, for example by permitting low-loss guidance of light in a hollow core, these fibers have important technological and scientific applications spanning many disciplines \cite{zolla_foundations_2005,Lourtioz_pc_2005,joannopoulos_photonic_2008,maier_plasmonics_2007}. The result has been a renaissance of interest in optical fibers, and their uses, over the past two decades as reviewed in \cite{markos_hybrid2017,hu_hybrid2019}; this activity has also strongly influenced acoustics through the developing field of phononic crystals \cite{laude_phononic_2015}. 

\begin{figure}[htb!]
\includegraphics[width=11.00cm]{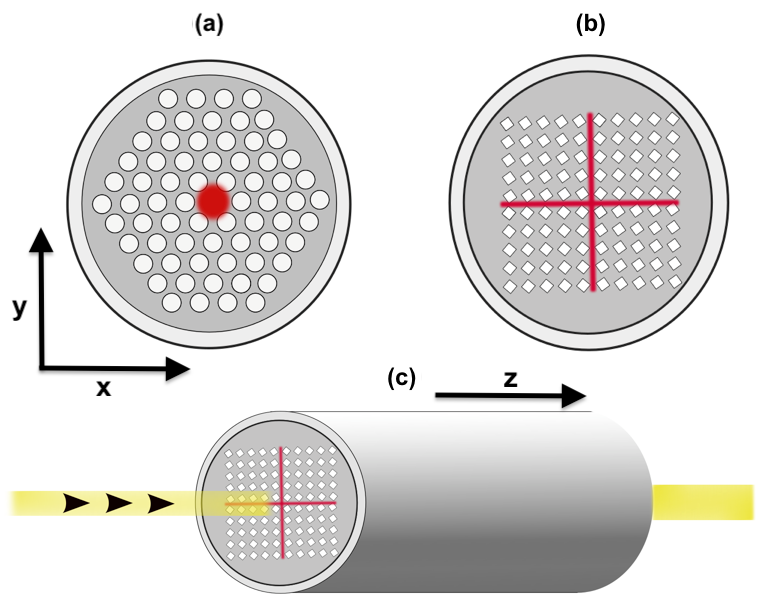}
\caption{Photonic band gap guidance versus hybrid topological-photonic guidance: (a) Conventional micro-structured fibre ($xy$-plane perspective) with light guided along a cavity defect surrounded by a crystal cladding; (b) Micro-structured fibre with light guided along interfaces between regions of oppositely orientated squares (c) 3D schematics of topological guidance; the red shows localisation of light, $z$-direction is also shown.}
\label{fig:intro_fig}
\end{figure}

Recently, ideas originating from topological insulators \cite{kane_z2_2005} have been transposed into photonics \cite{lu_topological_2014}, as reviewed in \cite{khanikaev_two-dimensional_2017}, showing promise for robust one-way edge states with enhanced protection against disorder. 
In Newtonian systems, this promise of topological protection is tempered by the requirement for time reversal symmetry (TRS) to be broken. An alternative, simpler approach, all be it resulting in less robust modes, is to leverage the pseudospins inherent within specific continuum systems. These pseudospin modes arise from strategically gapping Dirac cones to leave two pronounced valleys with locally quadratic curvature. This approach, commonly referred to as valleytronics \cite{xiao_valley-contrasting_2007, schaibley_valleytronics_2016, ren_single-valley_2015, ren_topological_2016}, is associated with the quantum valley-Hall effect in condensed matter physics. It has already been utilised in many settings, including the creation of dielectric photonic topological arrangements, leading to reflectionless guiding and designs for optical delay lines  \cite{ma_all-si_2016}.  Topological designs have also been recently implemented for telecommunication wavelengths \cite{shalaev_telecomms} on a CMOS-compatible chip thus bringing these concepts closer to application. These valley-Hall devices are locally topologically nontrivial however globally trivial, and therefore cannot draw upon the full power of the analogy with topological insulators, but they do have advantages in terms of their simplicity of construction as one needs  only to break spatial inversion or reflectional symmetries, together with actively suppressing backscattering. Symmetries are central to optical waveguide theory \cite{mcisaac_1975}, and have a long history within the field, and the topological guidance also draws upon symmetries and its mathematical language of group theory. Given the emergence of topological guiding there is now interest in developing this for photonic circuits, \cite{zhang_manipulation_2018}, and for Bragg fibres \cite{yeh78} by manipulating the cladding \cite{pilozzi_2020}, and there is a natural drive to explore the potential of these new ideas in other arenas. Our aim here is to demonstrate the viability of valleytronics to create localization of light for guidance in PCFs and to create a hybrid topological-PCF using these valley-Hall topological concepts (Fig. \ref{fig:intro_fig}). 

%Robustness re Prodan

 \begin{figure}[h!]
 \begin{center}
\includegraphics[width=11.0cm]{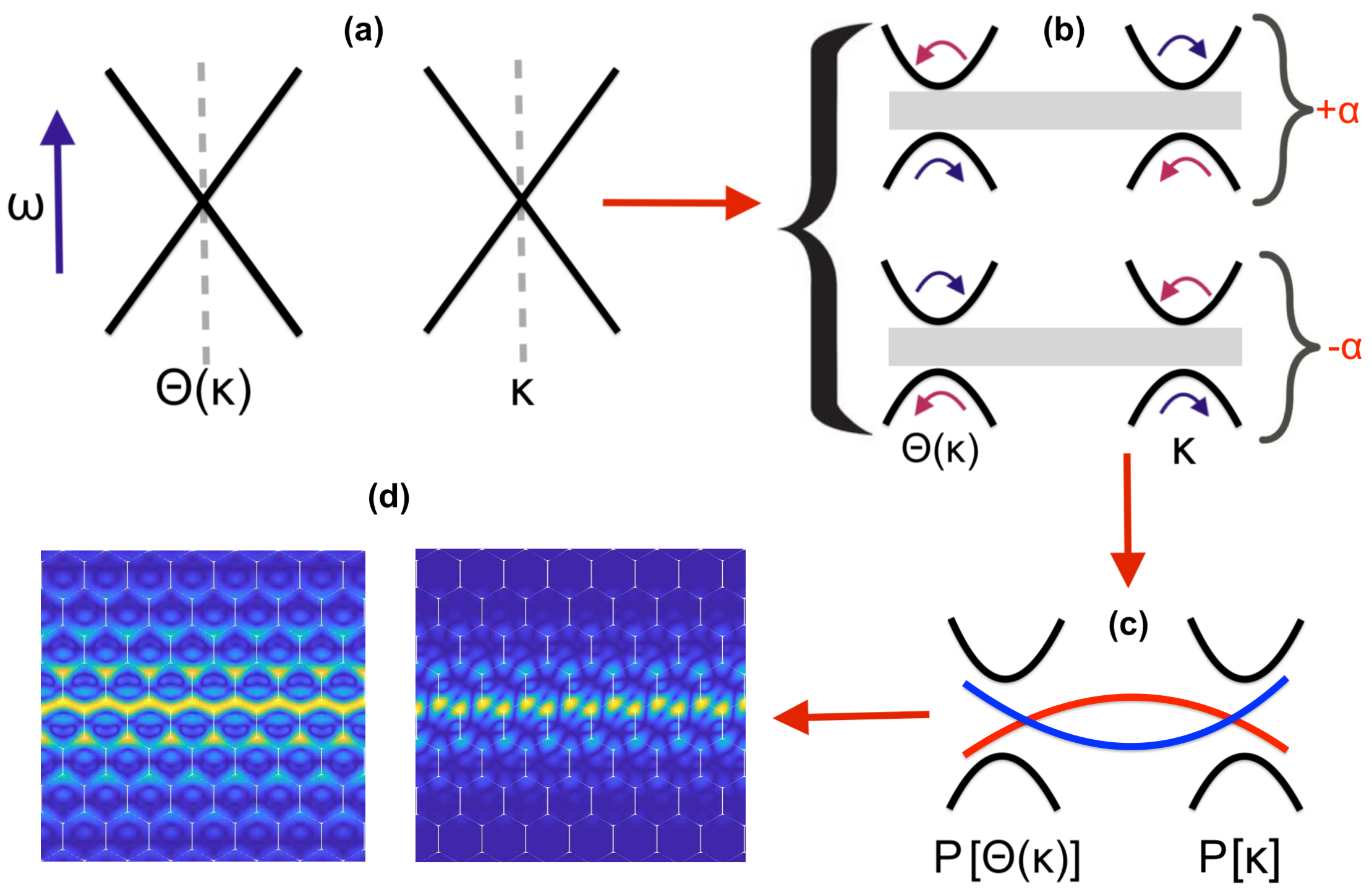}
\caption{Generation of topological modes: (a) Find two symmetry induced Dirac cones, well separated in Fourier space, and related by the TRS operator ($\Theta$), before applying a symmetry breaking perturbation, such as a rotation, $\pm \alpha$. (b) This perturbation gaps the Dirac cones leaving behind a set of 4 valleys that are locally imbued with a favoured chirality (as indicated by the arrows). The signum of the perturbation is directly related to the chiralities of valleys. (c) Subsequently we stack the $+\alpha$ medium above or below the $-\alpha$ medium to produce a pair of gapless edge states; the projection ($P$) of the Dirac cone pair marks the intersection point of the two curves. (d) Due to the underlying symmetry induced mechanism, these edge states are geometrically distinct; the edge states shown are typical, and illustrative, taken from the modelling in \cite{makwana_geometrically_2018}.}
\label{fig:topo_algo}
 \end{center}
\end{figure}

After summarising our problem formulation in Sec. \ref{sec:formulation} we begin in Sec. \ref{sec:symmetry}, by considering an infinite periodic photonic crystal with inclusions placed on a hexagonal or square lattice. The periodicity means that we can consider a single elementary cell and use Floquet-Bloch theory to generate an understanding of the system; dispersion curves that relate the frequency to Bloch wave-vector, portrayed around the edges of the Brillouin zone, encode the essential information about the system. 
We use our knowledge of symmetries in topological systems to obtain Dirac cones,  the chosen inclusions are then perturbed to gap the Dirac cones and thereby create band-gaps. Interfacial modes within the band-gaps, that benefit from topological concepts, are then used to develop the hybrid topological-PCFs. %The study proceeds in two steps. First, we consider Floquet-Bloch waves for PCFs with an infinite periodic cladding (for square and hexagonal lattices), and we analyse properties of the band diagrams and eigenfields.
 The steps to arrive at the localised edge-states, using topological ideas, are summarised in Fig. \ref{fig:topo_algo}. 
 These PCFs are investigated numerically, for finite array PCFs, in Sec. \ref{sec:topological_PCFs} and we use  finite element methods, and augment these numerics with a semi-analytic method based around induced monopole and dipole scattering by small cylinders suitable for open systems.

\section{Formulation}
\label{sec:formulation}

Although the harmonic Maxwell's equations (we assume $\exp({-i\omega t})$ dependence understood throughout) are inherently coupled, \cite{jackson_classical_1999}, provided one assumes infinite conducting boundary conditions, i.e. metallic fibers, along the fiber's axis \cite{compel_2004,zolla_foundations_2005}, the guided modes within a fiber decouple naturally into $p-$ and $s-$ polarizations. We consider an array of such fibers, see Fig. \ref{fig:intro_fig}, with periodicity in the $xy$-plane, and focus on the latter $s-$polarization. 
Taking into account the invariance of the fiber array along its $z$-axis, we look, for a given propagation constant $\gamma>0$, for
eigenfrequencies, $\omega$, and eigenmodes, ${\bf H}=\Re e({\boldsymbol{ \mathcal H}}(x,y)\exp(-i\gamma z))$, such that
 \begin{equation}
\nabla \times (\epsilon_r^{-1}\nabla\times {\bf H})=\mu_r\mu_0\epsilon_0\omega^2{\bf H}
\label{Maxwell}
\end{equation}
where $\nabla=(\partial/\partial_x,\partial/\partial_y,\partial/\partial_z)$, with $\epsilon_0,\mu_0$ ($\epsilon_r,\mu_r$) as the permittivity and permeability in-vacuum (and relative values); $\omega$ is angular frequency. % with time-harmonic waves, $\exp(-i\omega t)$, assumed.

In general, for oblique wave incidence down a PCF, we are faced with a fully-coupled vector electromagnetic problem \cite{zolla_foundations_2005} but, as noted above, in the case of crystal fibers with infinite conducting boundary conditions (a model for microwaves, or for light in dielectric fibers propagating with large propagation constant \cite{guenneau03a}), (\ref{Maxwell}) simplifies into
\begin{equation}
-\nabla_t \cdot (\epsilon_r^{-1}\nabla_t {H}_l)=\Omega^2 {H}_l
\label{Helmholtz}
\end{equation}
where $\nabla_t=(\partial/\partial_x,\partial/\partial_y)$,
$\Omega=\sqrt{\mu_r\mu_0\epsilon_0\omega^2-\gamma^2\epsilon_r^{-1}}$
and ${H}_l(x,y)$ is the longitudinal component of ${\boldsymbol{\mathcal H}}$. Eq. \eqref{Helmholtz} has the condition  $\epsilon_r^{-1}\nabla_t {H}_l\cdot{\bf n}=0$, on each fiber's boundary, where ${\bf n}$ is the unit outward normal to this boundary in the $xy$-plane.

An aside is that there is an equivalent phononic crystal fiber analogy: the same eigenvalue problem holds for time-harmonic scalar pressure acoustic waves $\Re e(P(x,y)\exp(-i\gamma z))$ propagating at an oblique angle $\gamma>0$ guided along an array of rigid fibers:
\begin{equation}
-\nabla_t \cdot (\rho_r^{-1}\nabla_t P)=\Omega^2 P
\label{HelmholtzPressure}
\end{equation}
 with $\rho_0,\kappa_0$ ($\rho_r,\kappa_r$) as the mass density and compressibility modulus in air (and relative values),
$\Omega=\sqrt{\kappa_r^{-1}\kappa_0^{-1}\rho_0\omega^2-\gamma^2\rho_r^{-1}}$. 

We naturally focus on PCFs, but note that our results hold for both time-harmonic electromagnetic waves in $s-$polarization, and pressure acoustic waves in phononics, propagating at an oblique angle within crystal fibers (with infinite conducting walls for the former, and rigid walls for the latter).
%{\color{blue}{one word of caution here for the reader would be better, the acoustic model is only valid for pressure waves propagating obliquely within a fluid, so air, water etc. so not a solid bulk medium otherwise there is coupling between p and s waves at the rigid inclusions, even in transverse incidence, which is actually what we do in the follow-up article.}}
%\subsection{Infinite crystal}
%\label{sec:infinite_crystal}
As we consider a periodic assembly of fibers, we use a normalised frequency, $\omega a/c$, hereafter,  where $a$ is the pitch of the lattice and $c$ the speed of light ($c^{-2}=\mu_0\epsilon_0\epsilon_r$ assuming $\mu_r=1$), or sound ($c^2=\kappa_0\kappa_r/(\rho_0\rho_r)$) in the medium surrounding the fibers.

%We checked some of our numerical results against existing benchmark results for PCFs obtained with the GETDP freeware \cite{dular_1998}.

\section{Symmetry protected topological states in hexagonal and square structures}
\label{sec:symmetry}

By considering an infinite array of identical parallel fibers we restrict our initial analysis to either a hexagonal or square elementary cell, containing a single or multiple fibers, that is then periodically repeated. For a given propagation constant $\gamma>0$, we find Floquet-Bloch waves such that
\begin{equation}
H_l({\bf r}+{\bf R}_p)=H_l({\bf r})\exp(i{\bf k}\cdot{\bf R}_p)
\label{Bloch}
\end{equation}
with ${\bf k}$ as the Bloch wavevector, ${\bf r}$ as the position vector and ${\bf R}_p=p_1{\bf a}_1+p_2{\bf a}_2$ as the vector attached to the nodes $(p_1,p_2)\in\mathbb{Z}^2$ of the lattice of translation vectors ${\bf a}_1$ and ${\bf a}_2$, which form the basis for the lattice as a whole; in this article we consider square and hexagonal lattices. 
 We proceed by numerically solving the spectral problem (\ref{Helmholtz})-(\ref{Bloch}) using the finite element package Comsol Multiphysics \cite{comsol} to compute the eigenfrequencies $\Omega_{\bf k}$ and associated Bloch eigensolutions ${H}_l^{\bf k}$ when ${\bf k}$ spans the Brillouin zone (BZ) in the reciprocal space. These finite element numerics are complemented by a semi-analytic method based around monopolar and dipolar scattering by small metallic cylinders, the details of which are in \cite{wiltshaw_neumann_2020} and covered briefly in Appendix \ref{sec:Neumann}. 
 
  These eigenfrequencies form a discrete spectrum $\omega_j$, $j=1, \, 2 \, \cdots$, and noting that $\Omega\geq 0$ in (\ref{Helmholtz}) it is easily
 seen  \cite{zolla_foundations_2005} that these are bounded below with 
% In general the waveguide is heterogeneous so the spectrum is 
 \begin{equation}
     {\gamma^2 c^2}%{\varepsilon_0\mu_0\sup_{(x,y)}\epsilon_r\mu_r}\leq \omega_1^2(\gamma)\leq\omega_2^2(\gamma)
     \leq\cdots \leq\omega_j^2(\gamma)\leq\cdots \; ,
 \end{equation}
and, as the propagation constant $\gamma$ varies, dispersion curves $\omega_j(\gamma)$ can be drawn.

%An empty discrete spectrum corresponds to frequencies under the cut-off, which in case of metallic PCFs with relative permeability $\mu_r=1$ and constant permittivity $\varepsilon_r$ is $\gamma/\sqrt{\epsilon_0\mu_0\epsilon_r}$. We note that if we let the radial outer boundary go to infinity, then modes are no longer bounded in the sense that they correspond to a complex
%frequency $\omega_j$. Indeed, in that case there is some energy leaking during the propagation
%of the signal and the imaginary part of the frequency corresponds to a decreasing exponential. The nickname of such modes is leaky modes \cite{white_2001}. Although it is not possible to capture such leaky modes with our model,

% We focus our attention in the sequel on eigenfields similar to topological edge states found in the periodic strips in section 3.1 and so associated with eigenvalues $\omega_j$ with numerical values very close to those of topological edge states, and we also show some typical defect modes to contrast the new type of hybrid topological modes with others.
 
 % The dispersion curves obtained for the hexagonal and square structure are shown in Figs. \ref{hex_bands}(c, d), \ref{sq_bands}(c, d), respectively.

\subsection{Hexagonal structures}
\begin{figure}[htb!]
\begin{center}
\includegraphics[width=10cm]{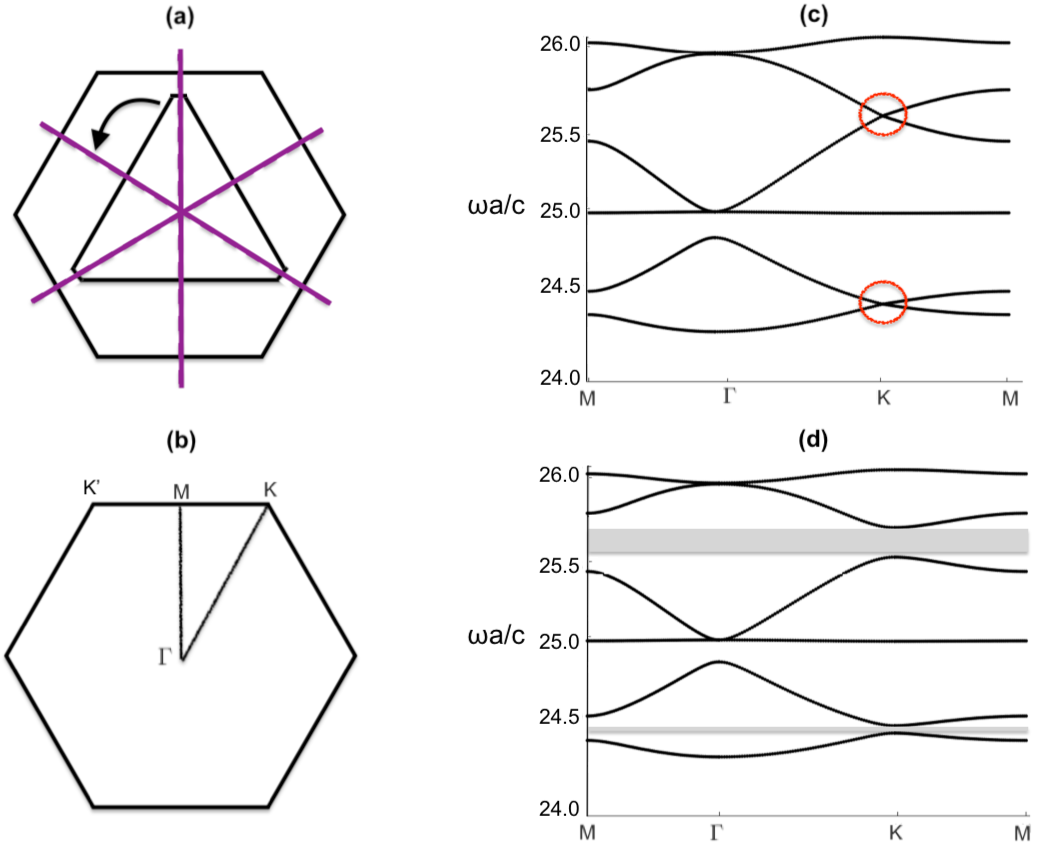}
\caption{PCF built from a hexagonal array, with lattice constant $a$, of a truncated triangular infinite conducting (or rigid in the phononic context) fibers, the elementary cell is shown in (a)  and contains an indented triangle defined by 3 long sides and 3 short sides of length $d_1= 1.65\, a$ and $d_2 = 0.11 \, a$ respectively, with the corresponding Brillouin zone in (b). Mirror symmetry lines are shown in (a) as dashed lines. 
Band diagrams for magnetic $H_l$ (or phononic pressure) eigenfield propagating at an oblique angle, characterised by (not normalized) propagation constant $\gamma=7$, and $a=2\sqrt 3$, 
%$\gamma a=7\times 2\sqrt{3}$,
 the unrotated case is shown in (c), and  when the fibers are tilted through an angle of 5 degrees about the center of the cell (d).
In (d) two topological band gaps open around the frequencies coincident with the circled Dirac points in (c).
%Creating a ribbon supercell made from tilted inclusions rotated clockwise above those rotated counter-clockwise leads to two interfacial edge waves  
The bandstructure in (c), that consists of a Dirac point and an entirely flat band, is reminiscent of the tight-binding honeycomb-Kagome structure shown in Fig. 14(a) of \cite{Barreteau_2017}.}
\label{fig:hex_bands}
\end{center}
\end{figure}

TRS topological-PCFs leverage the discrete valley degrees of freedom that arise from degenerate extrema in Fourier space (Fig. \ref{fig:topo_algo}). When constructing these PCFs we draw upon the valleytronics literature that, motivated by graphene, takes advantage of symmetry induced Dirac cones at the $KK^\prime$ vertices of the hexagonal BZ, see Fig. \ref{fig:hex_bands}(b, c) \cite{makwana_geometrically_2018, ochiai_photonic_2012}. These Dirac degeneracies are guaranteed if the structure belongs to one of three different symmetry sets \cite{makwana_geometrically_2018}: $\{G_{\Gamma}, G_{KK'}\} = \{C_{6v}, C_{3v}\}, \{C_{6}, C_{3}\}, \{C_{3v}, C_{3v}\}$, where $G_{\Gamma}, G_{KK'}$ denote the point group symmetries \cite{dresselhaus_group_2008, sakoda_book, inui_group_1990} 
 at $\Gamma$ and $KK'$, respectively.  We have chosen to illustrate the hexagonal topological PCFs using two of these cases $\{C_{3v}, C_{3v}\}$ and $\{C_{6}, C_{3}\}$. Turning to the  $\{C_{3v}, C_{3v}\}$ case first, shown in Fig. \ref{fig:hex_bands}(a), we gap the Dirac cone, and reduce the symmetry of the structure, down to $\{C_{3}, C_{3}\}$ by rotating the internal inclusion. The resulting band gap (Fig. \ref{fig:hex_bands}(b)) is demarcated by two regions of locally quadratic curvature commonly referred to as ``valleys". These valleys are imbued with a nonzero (local) topological quantity known as the valley Chern number; the opposite sign of the valley Chern numbers at the upper and lower valleys leads to the generation of topological edge modes \cite{qian_theory_2018, ochiai_photonic_2012}.  Importantly, the valleys have opposite chirality (or angular momenta) and are related by parity and/or reflectional symmetry as well as TRS. A clear demonstration in the context of photonics for delay lines is in \cite{ma_all-si_2016}, and for numerical convenience we adopted their truncated triangular inclusion. Despite this, we are not limited to using solely a triangular geometry for our fibre, as we may use any cellular structure that belongs to one of the three aforementioned Dirac cone inducing symmetry groups. To illustrate this further we also consider a geometry created from an arrangement of metallic cylinders. 

\begin{figure}[h!]
\includegraphics[width=14.00cm]{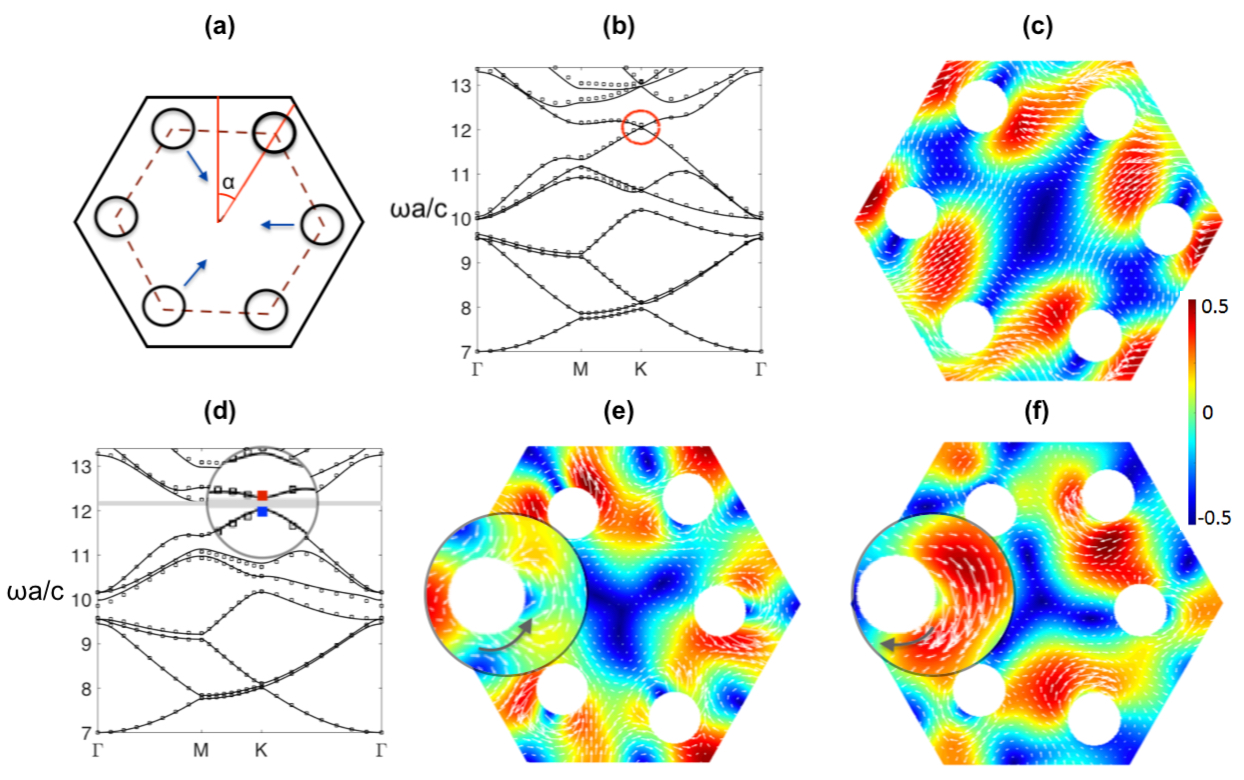}
\caption{PCF built from a hexagonal array  consisting of $6$ perfectly conducting cylinders of radius $\eta a = 0.07$ with  $\gamma = 7$ with lattice constant $a=1$. The elementary cell, shown in (a), has six-fold rotational symmetry and no reflectional symmetries; the centroid to scatterer distance is $0.400 a$, and the rotation angle $\alpha = \pi/6 + \pi/50$. The band diagram, shown in panel (b), is plotted around the irreducible Brillouin zone (IBZ) illustrated in Fig. \ref{fig:hex_bands}(b). An exemplar eigenfield associated with one of the degenerate eigenmodes is shown in (c). When three of the fibers are pulled towards the centroid of the cell and placed at a distance of $0.322 a$ from the centroid, as indicated by panel (a), we obtain the band diagram (d). The dispersion relations computed in (b) and (d) show agreement of results from finite element solutions (black lines) and scaled eigenvalues using the semi-analytic method from \cite{wiltshaw_neumann_2020}  (black squares) up to and including the bands of interest. The eigenstates that possess opposite chirality or angular momentum energy flux are shown in panels (e) and (f). These two eigenstates correspond to the lower (red) valley (e) and the upper (blue) valley (f) in panel (d).}
\label{fig:Neumann_bulk}

\end{figure}

\subsubsection*{Infinite array of small circular perfectly conducting cylinders}
We demonstrate how a different hexagonal geometry, Fig. \ref{fig:Neumann_bulk}(a), that belongs to a different symmetry set and uses circular inclusions, 
%\textcolor{red}{(in \cite{movchan_2007}, a PCF with similar geometry  to that in Fig. \ref{fig:Neumann_bulk} was analysed, but without any symmetry breaking, and some nearly flat bands were found similar to that in Fig. \ref{fig:hex_bands})} So what? Seems pretty tangential to me. 
 yields a topologically nontrivial band gap. 
The fabrication of circular fibres is easier than that of polygonal inclusions, and to emphasise the generality of the topological concepts we use circular inclusions arranged around a hexagonal perimeter. 
 %Not necessarily true. However, clearly easier to achieve than the Shvets triangles. That was the point.}
%\iffalse 
%In many practical cases, PCFs consist of a silica matrix with circular air-holes \cite{russell_2003,knight_2003,zolla_foundations_2005}
%For example, the fibers first demonstrated by Philip Russell consisted of a hexagonal lattice of air holes in a silica fiber, with a solid (1996) or hollow (1998) core at the center where light is guided. Other arrangements include concentric  rings of two or more materials, first proposed as "Bragg fibers" by Yeh and Yariv (1978), 
%\fi
 We opt to use an alternative semi-analytical  approach that is less computationally expensive than directly using finite element methods and which is relevant to open systems. This method is explained in detail in \cite{wiltshaw_neumann_2020} and is briefly summarised in Appendix \ref{sec:Neumann}. The usefulness of this semi-analytical scheme will become apparent in Sec. \ref{sec:topological_PCFs} when we design more complex topological PCFs.

% In this subsection, we not only use a different hexagonal geometry that yields topologically nontrivial band gaps, but also an alternative numerical approach that is less computationally expensive than FEM methods. This method is fully explained in (cite.) and is briefly summarised in Appendix \ref{sec:Neumann}. 

% The method (cite.) is reliant upon matched asymptotic expansions and is valid for small, but finite, Neumann scatterers. Using this system, (cite.) was able to derive a convergent numerical scheme based upon the truncated Fourier sum solution of the inhomogeneous Helmholtz equation. The ensuing matrix eigenvalue problem and its derivation is briefly summarised in Appendix \ref{sec:Neumann}. 

The geometry under consideration, Fig. \ref{fig:Neumann_bulk}(a), consists of several small, but finite radius, circular inclusions, located at the vertices of a hexagon. This smaller hexagon is contained entirely within the larger hexagonal cell and is rotated relative to it by an angle of $\theta = \alpha - \pi/6 = \pi/50$. This unperturbed arrangement belongs to the symmetry set $\{C_6, C_3\}$ and produces the dispersion curves shown in Fig. \ref{fig:Neumann_bulk}(b). By strategically perturbing the cellular structure (see arrows in Fig. \ref{fig:Neumann_bulk}(c)) the $C_6$ point group symmetry at $\Gamma$ is reduced down to $C_3$. This perturbation gaps the former Dirac cone to yield the band gap shown in Fig. \ref{fig:Neumann_bulk}(c) and we use the recipe outlined in Fig. \ref{fig:topo_algo}. The eigenstates, that sit at the blue and red marked valleys in Fig. \ref{fig:Neumann_bulk}(c), are illustrated in Fig. \ref{fig:Neumann_bulk}(e, f). The chirality of these eigenstates are characterised by the time-averaged energy flux, defined as,
\begin{equation}
\langle {\bf S} \rangle = H^*_l({\bf r}) \nabla_{\bf k} H_l({\bf r}),
\label{eq:flux}\end{equation} where $*$ denotes the complex conjugate. The magnifications in Figs. \ref{fig:Neumann_bulk}(e, f) demonstrates the opposite chirality (or angular momenta) between the pair of eigenstates. This angular momenta difference is a foundational property that predicates the generation of the topological edge states in Sec. \ref{sec:1D_dispersion}.

\subsection{Square structure}

Symmetry protected Dirac cones can also be constructed for non-hexagonal systems \cite{makwana_tunable_2019, makwana_topological_2019, Makwana_Experimental_2020, Jungmin_2020, xia_robust_2019}. Notably a strategically designed square (or rectangular) structure allows for the emergence of mirror symmetry protected Dirac cones; these systems differ from the vast majority of earlier valleytronic literature \cite{he_acoustic_2016, schomerus_helical_2010, ye_observation_2017, cheng_robust_2016, wu_direct_2017, xia_topological_2017, qiao_electronic_2011, makwana_geometrically_2018, tang_observations_2019} that have focused on graphene-like structures.

\begin{figure}[h!]
\begin{center}
\includegraphics[width=10cm]{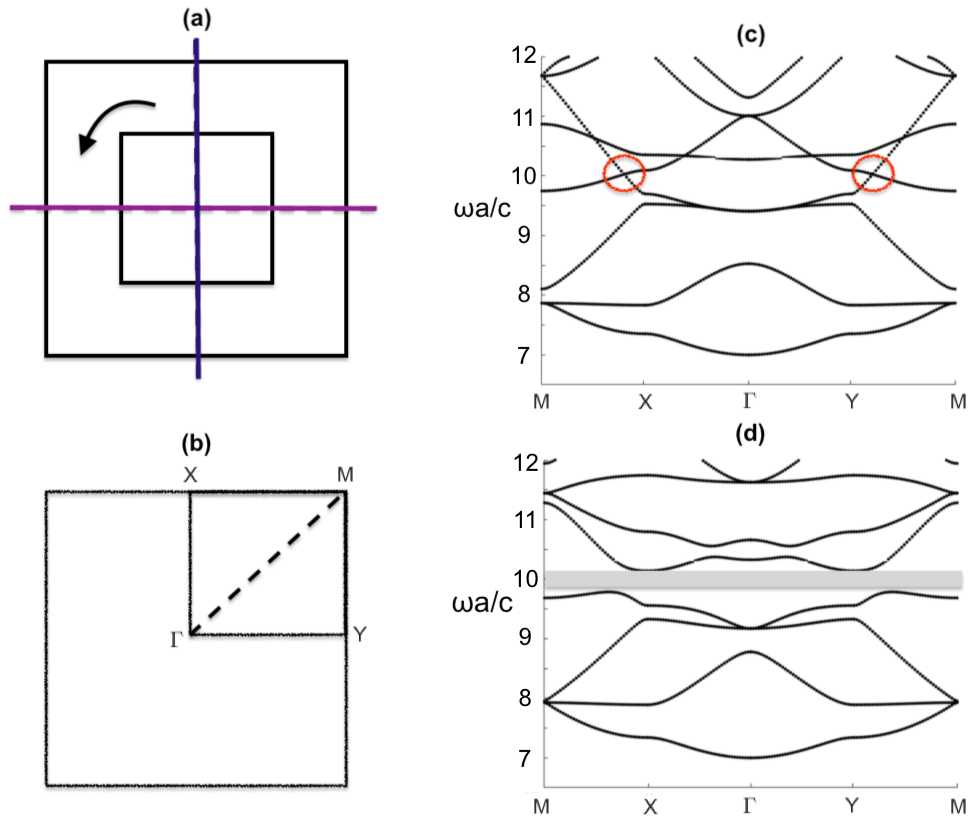}
\caption{PCF  built from a square array, with lattice constant $a$, of square infinite conducting (or rigid in the phononic context) fibers of cross-section $a^2/4$, the elementary cell is shown in (a) with the corresponding Brillouin zone in (b). 
Band diagrams for magnetic $H_l$ (or phononic pressure) eigenfield propagating at an oblique angle, characterised by (not normalized)  propagation constant, 
$\gamma=7$, $a=1$,
%$\gamma a=7$,
the unrotated case is shown in (c), and when fibers are tilted through an angle of 25 degrees about the center of the cell in (d).
In (d) a topological band gap opens around normalized frequency $\omega a/c=10$ coincident with the two circled Dirac points in (a).
%; within this stop band two interfacial edge waves exist (one odd, one even, shown on right hand side) when one considers a supercell with inclusions rotated clockwise and counter-clockwise.
%These edge waves propagate along the crystal fiber axis. {\color{blue}{All parameters have been normalized, vertical axis is $\omega a/c$, where $c$ is the wavespeed of light or sound in the medium surrounding fibers. DONE FOR THIS FIGURE.}
}
\label{fig:sq_bands}
\end{center}
\end{figure}

The unrotated cellular structure we choose, shown in Fig. \ref{fig:sq_bands}(a), contains, both, horizontal and vertical mirror symmetries along with four-fold rotational symmetry. These mirror symmetry lines are responsible for the highlighted Dirac cones located along the paths $MX$ and $MY$ in Figs. \ref{fig:sq_bands}(b, c) \cite{makwana_tunable_2019}. Unlike hexagonal structures, the position of these degeneracies can be tuned by varying the geometrical or material parameters of the system \cite{makwana_tunable_2019}. Despite the IBZ for this structure being an eighth of the BZ,  rather than a quarter, we plot around a quadrant of the BZ, Fig. \ref{fig:sq_bands}(b), as this incorporates the two distinct Dirac cones that are essential for our topological states. The desired quadrant has the following vertices: $X=(\pi/a,\pi/a)$, $N=(\pi/a,0)$, $\Gamma=(0,0)$, $M=(0,\pi/a)$, where $a$ is the lattice constant. The dispersion diagram, plotted around the IBZ, as well as a discussion about the influence of the diagonal mirror symmetries is in Appendix \ref{sec:Dihedral}.

The breaking of \emph{all} sets of mirror symmetries, via the rotation of the internal square fiber, leads to the topologically nontrivial band gap shown in Fig. \ref{fig:sq_bands}(c). The two residual valleys along each of the vertical and horizontal edges of the BZ are imbued with opposite sign valley Chern numbers. This difference in sign physically manifests itself as a difference in chirality between the valley eigenstates; this crucial property will be used in the next subsection.

 \begin{figure}[ht!]
 \begin{center}
\includegraphics[width=14.0cm]{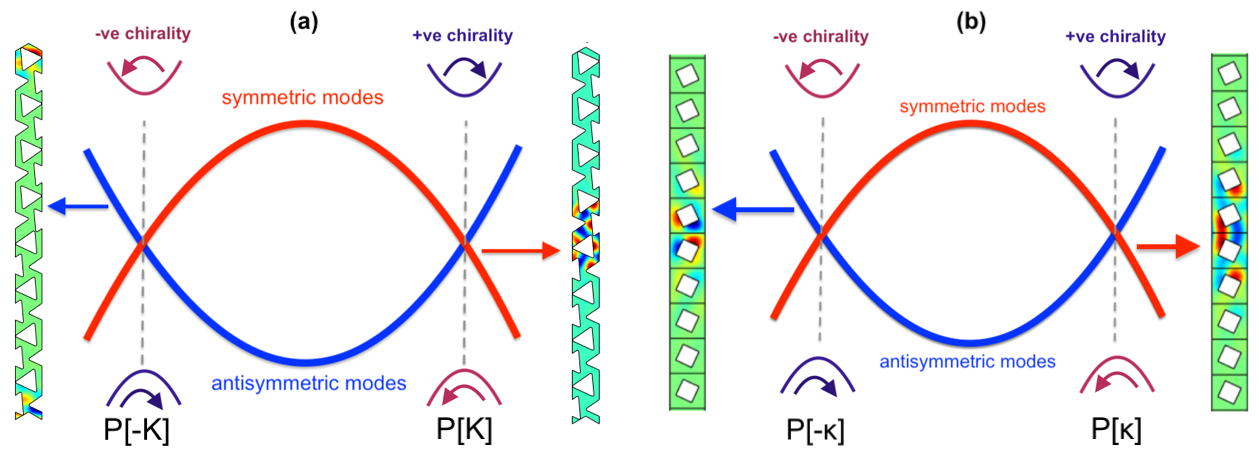}
\caption{(a) Schematic of the band diagram for a ribbon constructed from one medium stacked above the other (and with periodic boundary conditions between the top and bottom cells)
 such that the upper/lower cells contain fibers rotated clockwise/anti-clockwise (see Fig. \ref{fig:hex_bands}). The real parts of left/right ribbon eigenstates are shown evaluated at $\omega a/c = 25.36/25.66$. %\textcolor{red}{note that there is no unit here(kHz)} 
  (b) The analogous figure for the square structure shown in Fig.  \ref{fig:sq_bands}; frequencies of the left/right orthogonal eigenstates shown are $\omega a/c = 10.03/9.78$. %\textcolor{red}{no unit here(kHz)}.
 Unlike for the hexagonal case, the square's ribbon eigenmodes belong to the same interface rather than different interfaces.}
\label{fig:edge_modes}
 \end{center}
\end{figure}

\subsection{1D dispersion curves and edge states}
\label{sec:1D_dispersion}
The perturbation of the bulk media has broken reflectional or rotational symmetries whilst still retaining three-fold symmetry (for the hexagonal cases) and four-fold symmetry (for the square case). By stacking oppositely perturbed media, as shown in Fig. \ref{fig:edge_modes}, one above the other, we are able to create two seemingly distinct interfaces upon which edge states reside;  due to the broken symmetries, we now have inequivalent valleys characterised by opposite valley Chern numbers. Exemplar eigenstates, for the bulk media, located near the two topologically distinct valleys are shown in Fig. \ref{fig:Neumann_bulk}(e, f) and notably there is a marked difference between their angular momenta in prescribed regions. Creating interfaces, as we do here, between two media that have these opposite chiralities (or angular momenta) leads to the generation of topological edge states \cite{ochiai_photonic_2012, xiao_valley-contrasting_2007}; these are aptly named zero-line modes (ZLMs) due to the opposite sign of the valley Chern numbers on either side of the interface. The asymmetry, either side of the interface, leads to highly confined edge states such as those shown in Fig. \ref{fig:edge_modes}. A benefit of these topological modes is that we have a priori knowledge of how to construct the two adjoining media (sharing a band gap) such that we are guaranteed broadband edge modes that are more impervious to disorder \cite{xia_observation_2018} than competing designs. There is a notable difference between the square and hexagonal ribbon eigenmodes (Fig. \ref{fig:edge_modes}); the different parity eigenmodes belong to different interfaces for the hexagonal case and the same effective interface for the square case. This implies that a right-propagating mode along one of the square interfaces is also a left-propagating mode on the other. This nuance leads to additional functionality for square structures, absent from hexagonal structures, such as the three-way partitioning of energy away from a nodal point and the propagation of topological modes around a $\pi/2$ bend. Further discussion of this difference in behaviour between the two geometries is found in \cite{makwana_tunable_2019, makwana_topological_2019}.

% Commentary on robustness, long-range disorder, dependant upon Fourier separation etc. coupling between KK' modes.
Due to the absence of orthogonal spin$-1/2$ particles and the presence of TRS, caution is needed when we engineer our structure, in order to minimise inadvertent scatter \cite{makwana_designing_2018, qian_theory_2018}. There are competing mechanisms: the band gap needs to be large enough to localise the edge states near the domain wall, whilst not being so large, that the oppositely propagating modes are no longer orthogonal \cite{Wong_Kagome_2020}. The band gap width tunes the robustness and this is enhanced by preserving the locally quadratic curvature in the vicinity of the gapped Dirac cones. The strength of the perturbation is directly related to the orthogonality of the counter-propagating edge states. The Fourier separation between the modes excited should also be enhanced to limit inadvertent scattering \cite{makwana_designing_2018, qian_theory_2018}. Dirac cones located away from high-symmetry points, such as those of the square structure, are tunable in their position. Here, the robustness is further enhanced by artificially increasing the Fourier separation between the Dirac points which, in turn, leads to an increased separation between the forwards and backwards propagating edge modes. 

Numerical studies that demonstrate the robustness of these topological states versus conventional designs for Newtonian systems are found in \cite{xia_observation_2018, Zhixia_2020, Orazbayev_2019, morpurgo_intervalley_2006}. Theoretical studies that demonstrate the existence of these modes under disorder are found in \cite{lee-thorp_photonic_2016, fefferman_bifurcations_2016}. These studies pertain to quantum topological systems where the transition across the domain wall is approximated as a smooth $\tanh$ function. By carefully engineering our system, using the parameters at our disposal, namely, the perturbation strength, the Fourier separation and the transition across the interface, the valley has the potential to become a highly-efficient carrier of information \cite{khanikaev_two-dimensional_2017}.

\section{Finite topological photonic crystal fiber arrays}
\label{sec:topological_PCFs}

 \begin{figure}[htb!]
 \begin{center}
\includegraphics[width=14cm]{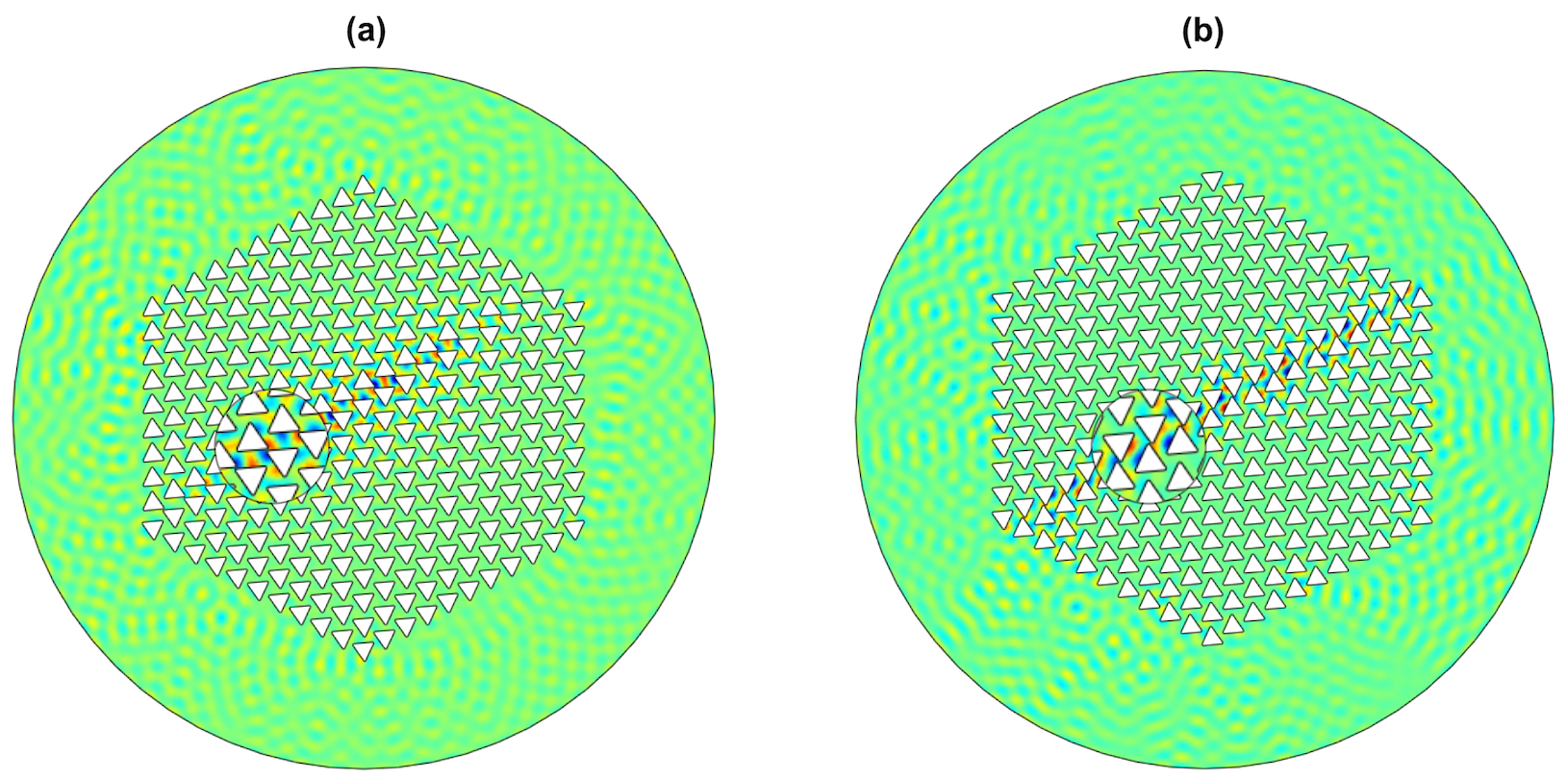}
\caption{Hexagonal ZLM: A finite PCF array constructed from 310 hexagonal cells, with two different media stacked over each other to create an interface, 
and outer boundary has radius $50$;  $\omega a/c=25.53, 25.54$ in (a,b) respectively and these frequencies lie in a topological band gap c.f. Fig. \ref{fig:hex_bands} (d).
 (a) has the upper/lower medium perturbed by a rotation clockwise/anti-clockwise and (b) perturbed by a rotation anti-clockwise/clockwise and each of these interfaces supports a distinct ZLM, c.f. Fig. \ref{fig:edge_modes} (a), as shown in the magnified regions. These numerical simulations utilise finite element methods and are for a closed system.   
}
\label{fig:Hex_ZLM}
 \end{center}
\end{figure}

 We now consider finite photonic crystal arrays created from metallic inclusions arranged to illustrate the topological edge states. The inclusions are arranged as a finite cluster, spaced with lattice constant $a$, and numerically we take the cluster to be surrounded by cladding.

 \begin{figure}[htb!]
 \begin{center}
\includegraphics[width=14cm]{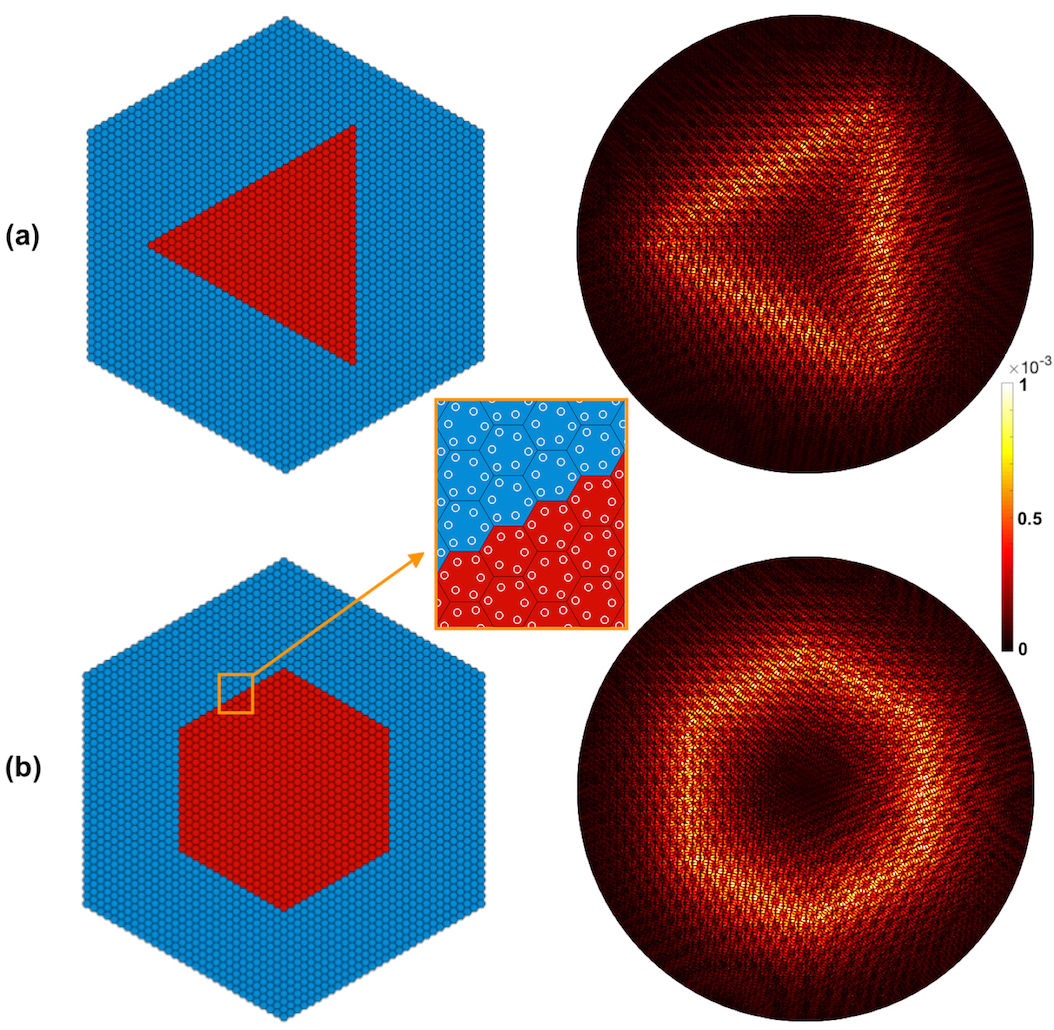}
\caption{ZLM guidance along more complex geometries taking a polygon constructed from one perturbed medium (red, created from $496$ and $721$ cells for (a) and (b) respectively) surrounded by an outer (blue) medium, with the opposite perturbation, see the inset for detail; we use the cell geometry of Fig. \ref{fig:Neumann_bulk}. The total structure in both (a) and (b) is formed from $2611$ cells hence a total of $15666$ circular scatterers. Panels (a) and (b) represent triangular and hexagonal polygonal patches; the red cell represents the arrangement given  Fig. \ref{fig:Neumann_bulk}(e,f), the blue cell is a $\pi/3$ rotation of the red cell. Each edge of the polygonal path supports a ZLM, and together they generate a ZLM along the whole boundary;  numerical simulations extract the localised states as shown and these occur at  frequencies $\omega a/c = 12.1422$ and $\omega a/c = 12.1213$ respectively for (a) and (b). 
%See Fig. \ref{SVDofFOLDY} for the homogenous Foldy solutions in which non-trivial modes are extracted without the requirement of an incident source.
 These simulations use the generalised Foldy method described in Appendix \ref{sec:Neumann} and are for an open system. 
}
\label{fig:neumann_pcf_2}
 \end{center}
\end{figure}

 We employ two different numerical methods, primarily we use the Finite Element package COMSOL 3.5 Multiphysics, implement equation (\ref{Helmholtz}) in the PDE module, and solve an  eigenvalue problems with the solver for Hermitian matrices (Lanczos algorithm). For these finite element simulations we have a radial outer boundary upon which we set metallic boundary conditions. It is well-known,  \cite{zolla_foundations_2005},  
  that this metallic outer boundary allows us to estimate the real part of the eigenfrequencies of an open system, provided the outer radial boundary is sufficiently large; we systematically checked that numerical values of eigenfrequencies of modes of interest are robust against the diameter of PCFs. To further confirm that these modes persist in the case of open PCFs we also utilise a semi-analytic method, designed for the rapid simulation of small metallic cylinders that asymptotically represents them as monopole and dipole scatterers  as detailed in \cite{wiltshaw_neumann_2020} and these simulations have no outer metallic boundary.

  We begin with an exploration of the ZLMs, for hexagonal media, containing the simple triangular inclusion of Fig. \ref{fig:hex_bands}, and take a hexagonal finite PCF as shown in Fig. \ref{fig:Hex_ZLM}. We cut this finite hexagonal array in half and rotate the inclusions in the upper/lower halves in opposite directions to form two distinct interfaces in a similar manner to Fig. \ref{fig:edge_modes}(a); these interfaces support a pair of geometrically distinct ZLMs \cite{makwana_geometrically_2018}. As noted in Sec. \ref{sec:1D_dispersion}, for the hexagonal system the relative ordering of the medium matters and we witness this from the edge states found for Fig. \ref{fig:Hex_ZLM}(a) and (b), which resemble those eigenmodes in Fig. \ref{fig:edge_modes}(a); the frequencies at which these edge states occur are reassuringly in the band gap predicted by the perturbed bulk media in Fig. \ref{fig:hex_bands}(d). These simulations are all done with standard finite element methods, i.e. Comsol, and the finite PCF array is surrounded by cladding and an outer metallic radial boundary thereby forming a closed system.

 \begin{figure}[b!]
  \begin{center}
\includegraphics[width=14cm]{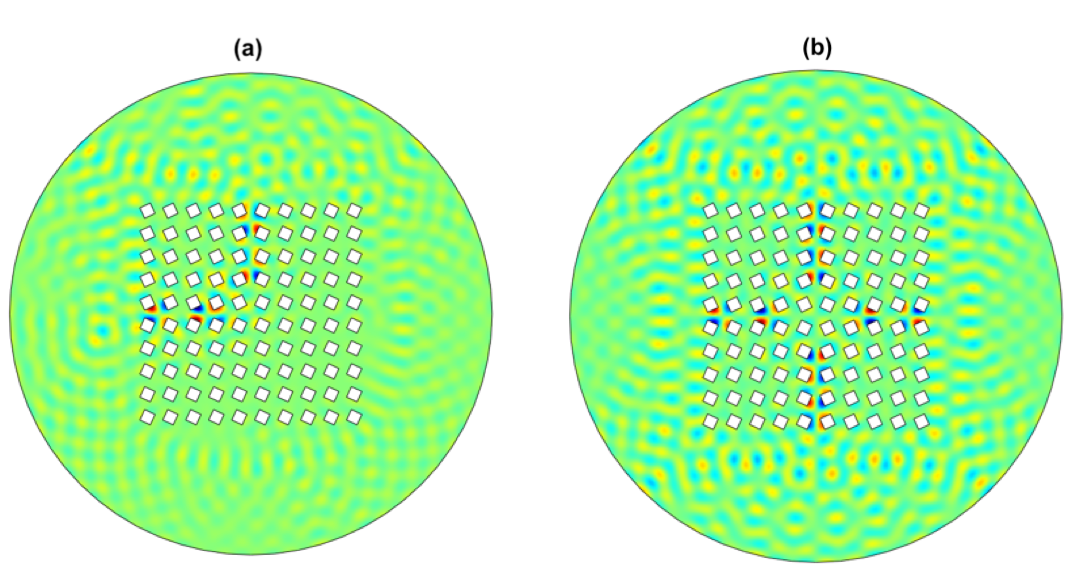}
\caption{Localisation in square PCFs: 
(a) a localised hybrid topological state confined to the edge of a quadrant 
%{\color{red} 98.26 Hz}, 
(d) a localised hybrid topological state confined by the interfaces of four quadrants % {\color{red} 98.62 Hz}.
Normalised frequencies are $\omega a/c= 9.54, 9.91$ in (a,b) respectively
 c.f. Fig. \ref{fig:edge_modes}(b) and the bulk topological band gap in Fig. \ref{fig:sq_bands}(d).
  These simulations used the finite element method. 
 }
\label{fig:square_chiral}
 \end{center}
\end{figure}

  The array in Fig. \ref{fig:Hex_ZLM} contains a relatively small number of scatterers, each of which is a single triangular shape, and the PCF is enclosed by a metallic outer boundary. We also want to consider more complex PCFs and explore whether we can generate topological states that remain localised around a \emph{closed} polygonal boundary with limited leakage. To do so,  we utilise the semi-analytical method, espoused in Appendix \ref{sec:Neumann}, to analyse complex open systems with ease.
  In Fig. \ref{fig:neumann_pcf_2} we consider two polygonal patches, a triangle and a hexagon, that are constructed from a set of adjoining zigzag interfaces \cite{xiao_valley-contrasting_2007} in the $xy$-plane; each of these crystals are formed from multiple smaller structured cells. The result of this construction are two striking highly localised triangular (Fig. \ref{fig:neumann_pcf_2}(a)) and hexagonal prism PCFs (Fig. \ref{fig:neumann_pcf_2}(b)); these fields consist of the topological modal conversion and preservation phenomena emphasised in \cite{makwana_geometrically_2018, tang_observations_2019}. Figs. \ref{fig:neumann_pcf_2}(a, b) take advantage of the bulk band diagrams of Fig. \ref{fig:Neumann_bulk}, and the construction methodology outlined in Fig. \ref{fig:topo_algo}. 
 
%\subsection{Hexagonal array of photonic crystal fibers}
%\label{sec:hexagonal}
%Same geometry as in \cite{ma_all-si_2016}
%Standard ZLM
%Vortex states paper \& refer to it etc, 

%\subsection{Square array of photonic crystal fibers}
%\label{sec:square}

The hexagonal array is the more common array type used in valleytronics, due to the guaranteed Dirac cones at the $KK'$ BZ vertices, but as shown earlier square arrays can also be used to create topological effects \cite{makwana_tunable_2019, makwana_topological_2019, Makwana_Experimental_2020, Jungmin_2020, xia_robust_2019}; we now construct small finite square arrays of just 100 square inclusions. We consider a circular metallic PCF of radius $10.5\, a$,
filled with $10\times 10$ square metallic rods of sidelength $a/2$ where $a$ is the array pitch. 
 The outer boundary at $10.5~a$ is taken to be metallic in simulations and we have checked that the modes we find are independent of radius and therefore are genuine trapped modes.

We use the geometry shown in Fig. \ref{fig:sq_bands}, and draw upon the edge states shown in Fig. \ref{fig:edge_modes}(b). Rather than just split a medium in half, we separate our square array into quadrants and then rotate the inclusions within each quadrant. Our motivation is to demonstrate localised topological modes, such as the $\pi/2$ bend and the four-armed cross, that are not possible when using the canonical hexagonal topological systems \cite{makwana_tunable_2019, makwana_topological_2019}. In Fig. \ref{fig:square_chiral} we illustrate localised states, along the edges of these quadrants, for this square array. Notably we rotate the square inclusions by 25 degrees either clockwise or anti-clockwise in different quadrants (similar to Fig. \ref{fig:edge_modes}(b)); panel (a) has the upper left quadrant rotated anti-clockwise and the remainder rotated clockwise, (b) has the upper left and lower right quadrants rotated anti-clockwise and the remainder rotated clockwise. As predicted from the topological theory \cite{makwana_tunable_2019, makwana_topological_2019} we expect localised hybrid topological states to form along the interfaces between the oppositely rotated inclusion quadrants and indeed this is observed in (a,b). There is, as also expected, a very slight frequency shift going between the predictions based around the infinite perfectly period crystal and numerical simulations based around the finite PCF of just $10\times 10$ rods nonetheless this is reassuring agreement. Although we have illustrated  just the odd mode in Fig. \ref{fig:edge_modes} there are complementary even mode solutions, see Appendix \ref{sec:Dihedral}, and indeed one can obtain combinations of these modes through cross-talk between the modes somewhat akin to that occurring in dual-core PCFs or between neighbouring open waveguides;     a linear combination of odd and even modes can be induced in neighbouring waveguides, and similarly for dual core PCFs, and we can obtain similar behaviour, but now between the horizontal and vertical interfaces. 
%also mention generation of supercontinuum, optical solitons etc. \cite{dudley06} in nonlinear PCFs.

\section{Conclusion}
\label{sec:conclusion}

We have introduced a new type of hybrid topological photonic/phononic band gap guidance that is topologically induced and utilised PCFs of metallic fibers, where a decoupling of Maxwell's equations occurs, to showcase the methodology and potential. Natural extensions involve moving to hollow or air-hole PCF, dielectric or hybrid PCFs and bringing topological effects into the fibre guidance of such more generalised systems. 
The construction of edge states using topological concepts is illustrated and follows a clear algorithm, c.f. Fig. \ref{fig:topo_algo}, and we generated interfacial edge waves propagating along the crystal fiber axis in two simple configurations, for a square array of square fibers and a hexagonal array of triangular fibers, in the case of longitudinal electromagnetic/acoustic pressure waves. To illustrate the generality of these concepts, and because cylindrical fibers are more commonly used,  these are augmented by a study using arrays of cylindrical inclusions placed at the vertices of a hexagon. 
Although here we have focused on infinitely conducting fibers, as this allows for a simple model, %valid for both microwaves and acoustic waves, 
 there are numerous extensions to draw upon in the topological valleytronics literature and beyond, and this will be a fruitful direction for future research. Examples include, but are not limited to, using valley vortex states or chirality locked beam-splitting \cite{ye_observation_2017, lu_valley_2016, wiltshaw_neumann_2020}, fragile topological states \cite{Paz_2019, Paz_2020}, non-Hermitian topological physics \cite{Zongping_2018} or even higher-order quadrupolar modes \cite{Benalcazar_2017}. 

In terms of physics, extensions to fully coupled electromagnetic waves in dielectric photonic crystal fibers and also for elastodynamic waves, as well as their interplay \cite{russell_sonic_2003} can be envisioned. Of particular interest is the case of open micro-structured waveguides, for which topological leaky modes would be associated with complex eigenfrequencies \cite{white_2001}, and thus would propagate over finite distances along the PCF axis \cite{kubota_2001}. It will be interesting to investigate whether they display less, or comparable, energy leakage to typical defect modes in hybrid, non topological, PCFs, especially with respect to a genuine (fibre drawing tower induced), or engineered, twist \cite{nicolet_2004,ma_2011}. 
% {\color{red} Other localised states induced using principles from topological physics?}
% Another interesting aspect of topologically guided hybrid modes is their robustness with respect to any defects in the cladding, which is valuable for fabrication purpose.

%Robustness Prodan...

%\textcolor{red}{we need to mention robustness of topological modes versus any defects in the cladding, this can be useful at the fabrication stage, I also suspect these modes should have less leakage than other non topological modes but this would require the open waveguide model.}

%We expect that our%
%work will foster theoretical and experimental efforts in these directions.

% {\color{red}Mehul: ensure that you mention fragile topological states, higher-order quadrupolar corner states, nonsymmorphic crystals, the obstructed atomic limit (i.e. Wu, Hu), topological quasicrystals, helically twisted Floquet photonic crystal fibres (sim. to Proceeding's paper) and any other relevant topological modes that would lead to confinement.}

%\textcolor{red}{For topological quasicrystals, you can refer to \cite{cai16} for a nice paper on quasicrystal fibres.}

\section*{Acknowledgments}
 MM and RC thank the EPSRC for their support through grant
{EP/L024926/1} and EP/T002654/1 and the support of European Union H2020 FETOpen project BOHEME under grant agreement No 863179.

%%%%%%%%%% If using BibTeX:
% \bibliography{sample, CompleteReferences_4}
% \bibliography{toto}

%%%%%%%%%% If preparing manually:
% \newcommand{\enquote}[1]{``#1''}

% \bibitem{Zhang:14}
% Y.~Zhang, S.~Qiao, L.~Sun, Q.~W. Shi, W.~Huang, L.~Li, and Z.~Yang,
%   \enquote{Photoinduced active terahertz metamaterials with nanostructured
%   vanadium dioxide film deposited by sol-gel method,}
%   {\protect\JournalTitle{Optics Express}} \textbf{22}, 11070--11078 (2014).

% {Optical Society}, \enquote{{OSA Publishing},}
%   \url{http://www.osapublishing.org}.

% \bibitem{FORSTER2007}
% P.~Forster, V.~Ramaswamy, P.~Artaxo, T.~Bernsten, R.~Betts, D.~Fahey,
%   J.~Haywood, J.~Lean, D.~Lowe, G.~Myhre, J.~Nganga, R.~Prinn, G.~Raga,
%   M.~Schulz, and R.~V. Dorland, \enquote{Changes in atmospheric consituents and
%   in radiative forcing,} in \enquote{Climate Change 2007: The Physical Science
%   Basis. Contribution of Working Group 1 to the Fourth assesment report of
%   Intergovernmental Panel on Climate Change,}  S.~Solomon, D.~Qin, M.~Manning,
%   Z.~Chen, M.~Marquis, K.~B. Averyt, M.~Tignor, and H.~L. Miler, eds.
%   (Cambridge University Press, 2007).

% \end{thebibliography}
\appendix

\section{Planar array of small circular scatterers}
\label{sec:Neumann}

% \subsection{Matrix eigenvalue problem}
Equations \eqref{Helmholtz} and \eqref{HelmholtzPressure} are, for metallic scatterers and with $\epsilon_r$ constant, the Helmholtz equation and we can draw upon efficient, rapid, semi-analytical methods created for small inclusions as described in \cite{martin2006multiple}. These are derived in \cite{wiltshaw_neumann_2020} (based upon an extension of \cite{schnitzer2017bloch}) utilizing matched asymptotic expansions, each circular scatterer - of radius $\eta a$ - is approximated by a monopole and dipole source term whose coefficients, $a_{\mathrm{mono}}$ and $\textbf{b}_{\mathrm{di}}$ are \textit{a priori}  unknown constants. These unknown constants can trivially absorb factors of $a$ or $\epsilon_{r}$, the analysis from \cite{wiltshaw_neumann_2020} remains largely unchanged, that is for a single scatterer
\begin{equation}
     \left( \nabla^{2}_{t} + \epsilon_{r} \Omega^{2} \right) H_{l} = 4i \eta^{2} \left\lbrace a_{\mathrm{mono}} - \textbf{b}_{\mathrm{di}} \cdot \nabla_{t} \right\rbrace \delta(x,y). \label{HelmholtzNeumannApprox}
\end{equation}
The assumptions being that the scatterers are small, such that any defined inner region is asymptotically valid; moreover, the analysis requires that the scatterers are not too closely spaced such that any matched asymptotic analysis breaks down. 

As shown in \cite{wiltshaw_neumann_2020} this approach leads to a matrix eigenvalue problem that gives the Floquet-Bloch dispersion curves, this is effectively a variant upon the plane wave expansion method \cite{johnson_block-iterative_2001}  but specialised to small inclusions. This approach can also be generalised to do scattering simulations efficiently for very large numbers of inclusions as an extension \cite{martin2006multiple} of Foldy's method \cite{foldy1945multiple} which is often used for small monopolar scatterers. 

%for more information regarding the matrices within these schemes; wherein we discuss the homogeneous Foldy problem, in which if we set $\mathrm{B}_{\mathrm{Foldy}} = \textbf{0}$ 

In the main text, see Fig. \ref{fig:neumann_pcf_2} we extract localised modes around polygonal arrangements of many inclusions arranged either in a triangle or hexagon; technically we pull out these frequency dependent non-trivial modes %which will propagate throughout the finite structure - provided we can find a non-trivial solution. The best non-trivial solution at our disposal is the
using singular value decomposition.

\section{Effects of the diagonal mirror symmetry line on square structures}
\label{sec:Dihedral}
\begin{figure}[h!]
    \centering
    \includegraphics[width=14cm]{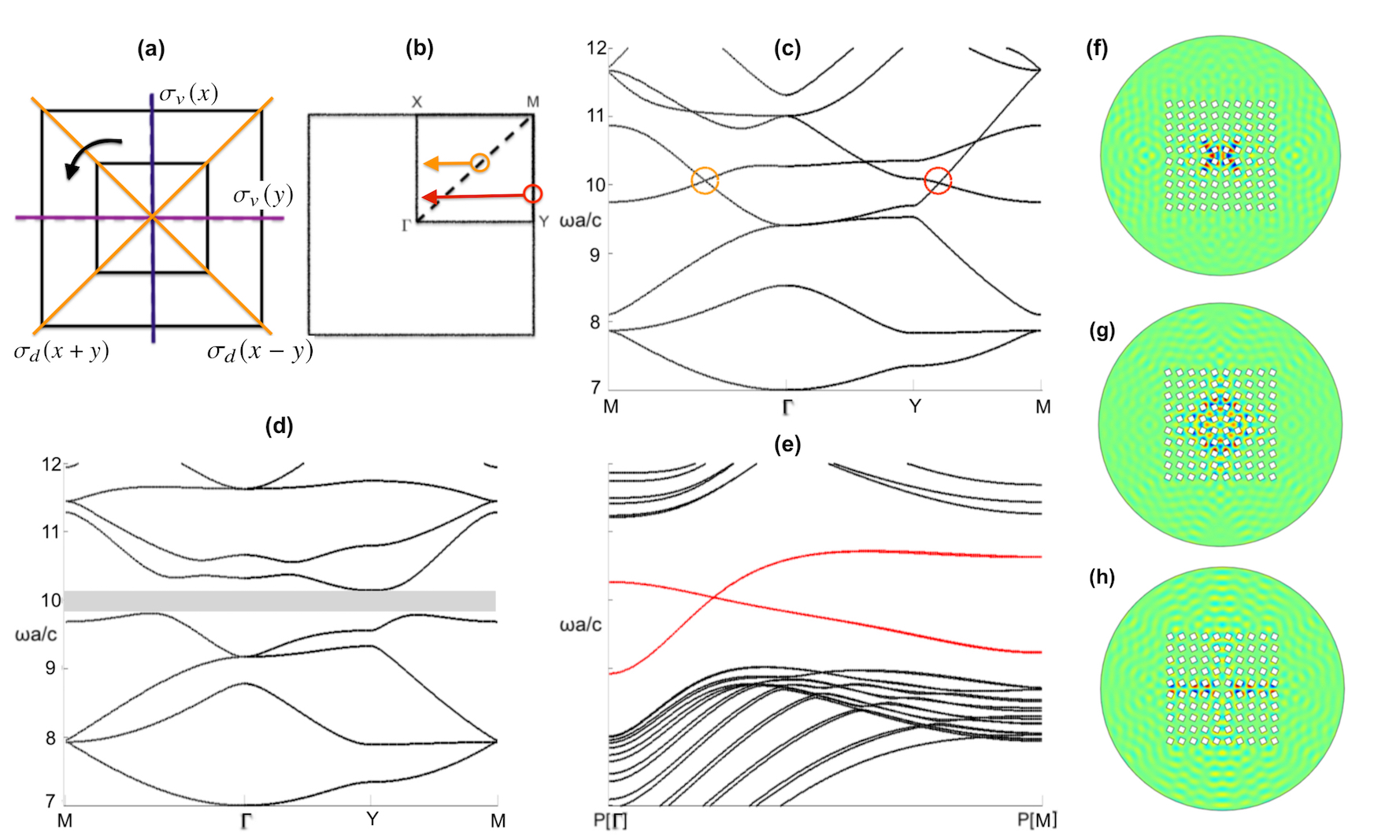}
    \caption{Panel (a) shows the physical space elementary cell containing a square fibre; the structures reflectional symmetries $\sigma_v(x, y), \sigma_d(x \pm y)$ are also shown. The BZ, IBZ and quadrant are shown in (b) alongside the locations of the $\sigma_v(x, y), \sigma_d(x \pm y)$ induced Dirac cones (c), the arrows indicate their projection locations in the $1D$ bandstructures (e). The Dirac cones in (c) are gapped to yield (d). Other exemplar localised modes, associated with the square structure, are shown in panels (f) $\omega a/c = 10.144$, (g) $\omega a/c = 10.09$ and (h) $\omega a/c = 9.93$.}
    \label{fig:sigma_d}
\end{figure}

The point group symmetry of the unperturbed structures of Fig. \ref{fig:sq_bands}~(a) and Fig. \ref{fig:sigma_d}~(a) is $C_{4v}$; this arises from the four-fold rotational symmetry and the two sets of of mirror symmetry lines, $\sigma_v(x, y)$ and $\sigma_d(x \pm y)$. The vertical and horizontal mirror symmetries, $\sigma_v(x, y)$, yield the non-symmetry repelled Dirac cones along the outer perimeter of the BZ, see Fig. \ref{fig:sigma_d} (b, c) and \cite{makwana_tunable_2019}.  The particular Dirac cone shown in Fig. \ref{fig:sigma_d}(c) is dependent upon $G_M = C_{4v}$ and $G_{MY} = \sigma_v(x)$; due to the continuity of the bands, across the high-symmetry point $M$, the irreducible representations (IR) of the $C_{4v}$ group transform into those in the $C_s$ group \cite{dresselhaus_group_2008, inui_group_1990}. The two bands along $MY$, that eventually cross, have the IRs $A_1, B_2$ at $M$, these transform via the compatibility relations \cite{dresselhaus_group_2008, inui_group_1990} into the opposite parity $A, B$ IRs along $MY$. This implies that a careful parametric tuning of our system can lead to an unavoidable crossing between the bands \cite{makwana_tunable_2019} and this is precisely what occurs for our system, see Figs. \ref{fig:sq_bands}(c), \ref{fig:sigma_d}(c). A similar mechanism occurs, due to the $\sigma_d(x \pm y)$ symmetries, whereby the $A_1, B_2$ IRs transform into the $A, B$ IRs along the path $\Gamma M$ thereby leading to the non-symmetry repelled crossing shown in Fig. \ref{fig:sigma_d}(c). The rotation of the square fiber leads to the band gap shown in Fig. \ref{fig:sigma_d}(c). This band gap is demarcated by the valleys along $MY$ (and by those along $MX$), not by the $\sigma_d(x \pm y)$ induced valleys and hence the band gap width in Fig. \ref{fig:sigma_d}(d) matches that of Fig. \ref{fig:sq_bands}(d).

The computed edge states formed via the stacking of one perturbed medium above its reflectional twin, Fig. \ref{fig:edge_modes}(b), are shown in Fig. \ref{fig:sigma_d}(e). The gapless nature of the states is indicative of their underlying topological nature. For hexagonal systems, ZLMs with distinguishable valley degrees of freedom exist for every propagation direction other than for the armchair \cite{fefferman_bifurcations_2016}; the armchair termination superposes the $KK'$ valleys, thereby coupling them and making them less robust; this coupling results in band repulsion and gapped edge states \cite{heine_group_nodate}.

Other examples of highly confined edge states using four structured quadrants, Fig. \ref{fig:square_chiral}(b), are shown in Fig. \ref{fig:sigma_d}(f, g, h). The coupling between the even and odd-parity modes in these figures resembles the cross-talk phenomena found in multiple core optical waveguides \cite{yamashita_1985} and PCFs. 

\end{document}